\documentclass[twocolumn,superscriptaddress,aps,prl,amssymb,amsfonts,amsmath,floatfix]{revtex4-1}
\usepackage{bm}
\usepackage{xcolor}
\usepackage{hyperref}
\usepackage{graphicx}
\usepackage[normalem]{ulem}

\begin{document}

\title{Finite-disorder critical point in the yielding transition of elasto-plastic models}%

\author{Saverio Rossi}%
\affiliation{LPTMC, CNRS-UMR 7600, Sorbonne Universit\'e, 4 Pl. Jussieu, F-75005 Paris, France}

\author{Giulio Biroli}%
\affiliation{Laboratoire de Physique de l’Ecole Normale Sup\'erieure, ENS, Universit\'e PSL, CNRS, Sorbonne Universit\'e, Universit\'e Paris Cit\'e, F-75005 Paris, France}

\author{Misaki Ozawa}%
\affiliation{Laboratoire de Physique de l’Ecole Normale Sup\'erieure, ENS, Universit\'e PSL, CNRS, Sorbonne Universit\'e, Universit\'e Paris Cit\'e, F-75005 Paris, France}

\author{Gilles Tarjus}%
\affiliation{LPTMC, CNRS-UMR 7600, Sorbonne Universit\'e, 4 Pl. Jussieu, F-75005 Paris, France}

\author{Francesco Zamponi}%
\affiliation{Laboratoire de Physique de l’Ecole Normale Sup\'erieure, ENS, Universit\'e PSL, CNRS, Sorbonne Universit\'e, Universit\'e Paris Cit\'e, F-75005 Paris, France}

\date{\today}%

\begin{abstract}
Upon loading, amorphous solids can exhibit brittle yielding, with the abrupt formation of macroscopic shear bands leading to fracture, or ductile yielding, with a multitude of  plastic events leading to homogeneous flow.
It has been recently proposed, and subsequently questioned, that the two regimes are separated by a sharp critical point, as a function of some control parameter characterizing the intrinsic disorder strength and the degree of stability of the solid. In order to resolve this issue, we have performed  extensive numerical simulations of athermally driven elasto-plastic models with long-range and anisotropic realistic interaction kernels in two and three dimensions. Our results provide clear evidence for a finite-disorder critical point separating brittle and ductile yielding, and we provide an estimate of the critical exponents in 2D and 3D. 
\end{abstract}

\maketitle

Yielding of amorphous materials is a practically and scientifically important problem~\cite{nicolas2018deformation,barrat2011heterogeneities,rodney2011modeling,bonn2017yield,falk2011deformation}.
When a material is mechanically slowly driven from
an initial quiescent glassy state, two different types of
yielding behavior are observed. One is brittle yielding, where the sample catastrophically breaks into pieces
and displays one or several macroscopic shear bands (usually experimentally encountered in atomic and molecular glasses). The other one is
ductile yielding, for which the sample deforms rather homogeneously via a series of local and mesoscopic plastic events that prevent catastrophic failure (usually experimentally encountered in soft materials like colloids and pastes). It has been established that a given material may show brittle or ductile yielding depending on the preparation history of the sample~\cite{rodney2011modeling,kumar2013critical,vasoya2016notch,fan2017effects}. In particular,
a well-annealed, hence stable, glass sample  shows brittle yielding, whereas a poorly-annealed, less stable, glass
sample exhibits ductile yielding.
Note that in the materials science and engineering communities the yield point is traditionally  defined as the end of the purely elastic branch and the onset of plastic behavior. However, because several molecular simulations and elasto-plastic model (EPM) studies (see, e.g.,~\cite{karmakar-lerner-procacciaPRE82-2010,lemaitre2021anomalous,lin2015criticality}) demonstrated that a purely elastic branch does not exist in sheared amorphous solids as plasticity appears for any infinitesimal deformation in the thermodynamic limit, the statistical physics community usually adopts a different definition of yielding.

Recent theoretical studies suggest that brittle yielding corresponds to a nonequilibrium 
first-order transition (or spinodal~\cite{wisitsorasak2012strength,rainone2015following,urbani2017shear,jaiswal2016mechanical,parisi2017shear,de2020renormalization}), 
associated with a macroscopic discontinuous stress drop at a given strain value, whereas ductile yielding corresponds to a continuous stress-strain curve, corresponding to a progressive plastic softening of the material.
In athermal quasi-static (AQS) conditions~\cite{maloney2006amorphous}, 
it was observed that these two distinct behaviors are separated by a critical value of the stability (or the disorder)~\cite{ozawa2018random,ozawa2020role,Rossi_2022}. It was then proposed that the brittle-to-ductile transition is a novel nonequilibrium phase transition, similar to that of an  athermally driven random-field Ising model~\cite{sethna2001crackling,nandi2016spinodals}.
Further understanding the transformation from brittle to ductile yielding appears as a major challenge in many fields, from materials science to statistical physics~\cite{dauchot2011athermal,ketkaew2018mechanical,jin2018stability,yeh2020glass,bhaumik2021role,barbot2020rejuvenation,da2020rigidity,sastry2021models,benzi2021continuum,langer2020brittle,richard2021brittle,arceri2021marginal}.

The above scenario has been challenged in  Refs.~\cite{barlow2020ductile,PhysRevResearch.4.043037,richard2021finite}. In particular, Ref.~\cite{barlow2020ductile} argues that in  AQS condition and provided the samples are large enough, yielding always takes place in a brittle manner. This should happen irrespectively of the stability or disorder of the samples, except for the putative infinitely disordered sample, thereby implying that the brittle-to-ductile transition does not exist in the thermodynamic limit. Large-scale molecular simulations~\cite{richard2021finite} seem to give some support to the statements of Refs.~\cite{barlow2020ductile,PhysRevResearch.4.043037}. Yet, it remains hard to conclude due to the limited system sizes accessible in molecular simulations and the very small number of samples involved.

In this Letter, in order to overcome this difficulty and obtain conclusive results, we perform a thorough numerical analysis of the brittle-to-ductile transition in EPMs~\cite{picard2004elastic,nicolas2018deformation}. 
These mesoscopic models have already been successfully applied to describe  several aspects of the rheology of amorphous materials, in particular the yielding transition \cite{popovic2018elastoplastic,PhysRevResearch.4.043037}.  Their coarse-grained lattice nature enables us to access very large system sizes and a large number of samples, allowing for a careful finite-size scaling analysis of the critical point. Our main result is a direct numerical evidence for the existence of a finite-disorder critical point separating brittle and ductile behavior.

The type of EPM we focus on provides a simple scalar description of the AQS dynamics~\cite{baret2002extremal,vandembroucq2011mechanical,lin2014scaling} and corresponds to a cellular automaton on two-dimensional (2D) and three-dimensional (3D) cubic lattices. 
In particular, we consider incompressible, homogenous and isotropic materials under a simple shear deformation protocol and we focus on a single shear-stress component that we denote $\sigma$~\cite{picard2004elastic}. (This is an approximate treatment which ignores the other stress components.)
The model describes the evolution of coarse-grained local stresses $\sigma_i$ in the presence of an external strain $\gamma$. Whenever one such stress goes above a stability threshold, the site yields, and the resulting stress drop is propagated through the sample via a  long-range Eshelby-like propagator~\cite{eshelby1957determination}.
The initial stability of the solid, which in real systems depends on the annealing protocol, is quantified by a parameter $R$ associated with the width of the initial stress distribution (and therefore characterizing the strength of the disorder). 
We vary the system size over a wide range of linear box lengths, $L=256-4096$ for 2D and $L=48-164$ (with a few samples at $L=200$) for 3D. 
Details concerning the simulated model and the numerical simulations are presented in the SM  \cite{supp_arxiv}. 
We have checked that variations of the model corresponding to different ways of accounting for the initial stability and the force balance lead to the same results (see the SM). 

\begin{figure}
    \centering
    \includegraphics[width=\linewidth]{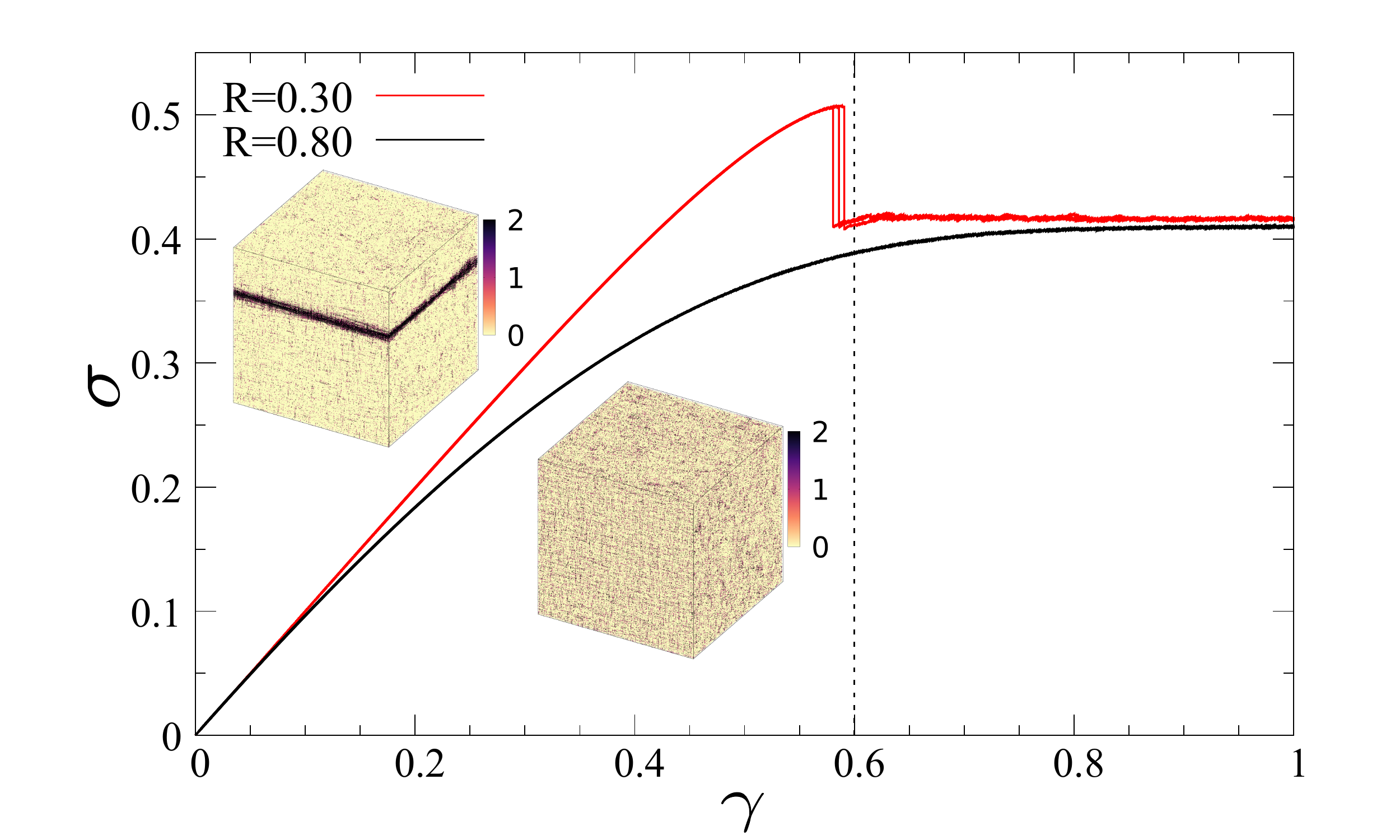}
    \caption{Stress versus strain curves in the 3D elasto-plastic model illustrating the brittle ($R=0.3$) and the ductile ($R=0.8$) cases. The linear box length is $L=200$. Three independent samples are presented. Insets: Real-space configurations at $\gamma=0.6$ in the two cases. The color bar corresponds to the number of local plastic events.
    }
    \label{fig:largesample_stress_vs_strain}
\end{figure}

We first show that the model displays the same behavior as that found in numerical simulations of particle systems \cite{rodney2011modeling,ozawa2018random,ozawa2020role}, with in particular the signature of a brittle-to-ductile transition accompanied by substantial finite-size effects \cite{PhysRevResearch.4.043037,richard2021finite}. In consequence, it provides a suitable framework to address the issues discussed above. In Fig.~\ref{fig:largesample_stress_vs_strain}, we present stress-versus-strain curves for a 3D system with $L=200$ for two values of the disorder strength, $R=0.3$ and $R=0.8$. The former clearly shows brittle yielding characterized by a discontinuous stress drop and the appearance of a shear band (top inset), while the latter displays ductile yielding characterized by a continuous monotonic stress growth and homogeneously distributed plastic events (bottom inset).
Brittle and ductile yieldings are thus qualitatively distinct, and their occurrence depends on the disorder strength $R$. The present results are in line with previous numerical observations in two-dimensional EPMs~\cite{vandembroucq2011mechanical,popovic2018elastoplastic,liu2022fate}.

Our second goal is to  identify the putative critical point separating brittle and ductile yielding.
In previous studies~\cite{ozawa2018random,ozawa2020role,bhaumik2021role}, the maximum stress drop, $\langle \Delta \sigma_{\rm max} \rangle = \langle \max_{\gamma}\{\Delta \sigma(\gamma)\}\rangle$  where $\Delta \sigma(\gamma)$ is the stress drop due to irreversible events at the strain $\gamma$, was used as an order parameter to detect the transition. Here, $\langle \cdots \rangle$ denotes an average over many independent realizations (or samples),
and the maximum is computed for each sample and then averaged over samples. We have found that for EPMs a more efficient order parameter is obtained from the fraction of sites along a single line in 2D or plane in 3D that have yielded at least once up to the strain $\gamma$. We consider the maximum of this fraction over all horizontal and vertical lines (2D) or planes (3D) and we call this quantity $n(\gamma)$. By construction it is an increasing function of $\gamma$. It shows a discontinuous jump of order O(1) when $\sigma(\gamma)$ shows a discontinuous drop of order O(1) and it increases continuously when $\sigma(\gamma)$ shows a continuous ductile behavior (see below). Therefore, $n(\gamma)$ essentially contains the same information as $\sigma(\gamma)$ for distinguishing brittle and ductile yielding behavior. We then define a new order parameter, $\Delta n_{\rm max} = \max_{\gamma}\{\Delta n(\gamma)\}$, where $\Delta n(\gamma)$ is the jump of $n(\gamma)$ that takes place in the AQS dynamics from $\gamma$ to $\gamma+\Delta \gamma$ in a given sample. We have observed that $\Delta n_{\rm max}$ better quantifies the abrupt emergence of a system-spanning shear band and, as a result, detects the critical point in EPMs more accurately than $\Delta \sigma_{\rm max}$ (see the detailed discussion in the SM). 

\begin{figure}
    \centering
    \includegraphics[width=\linewidth]{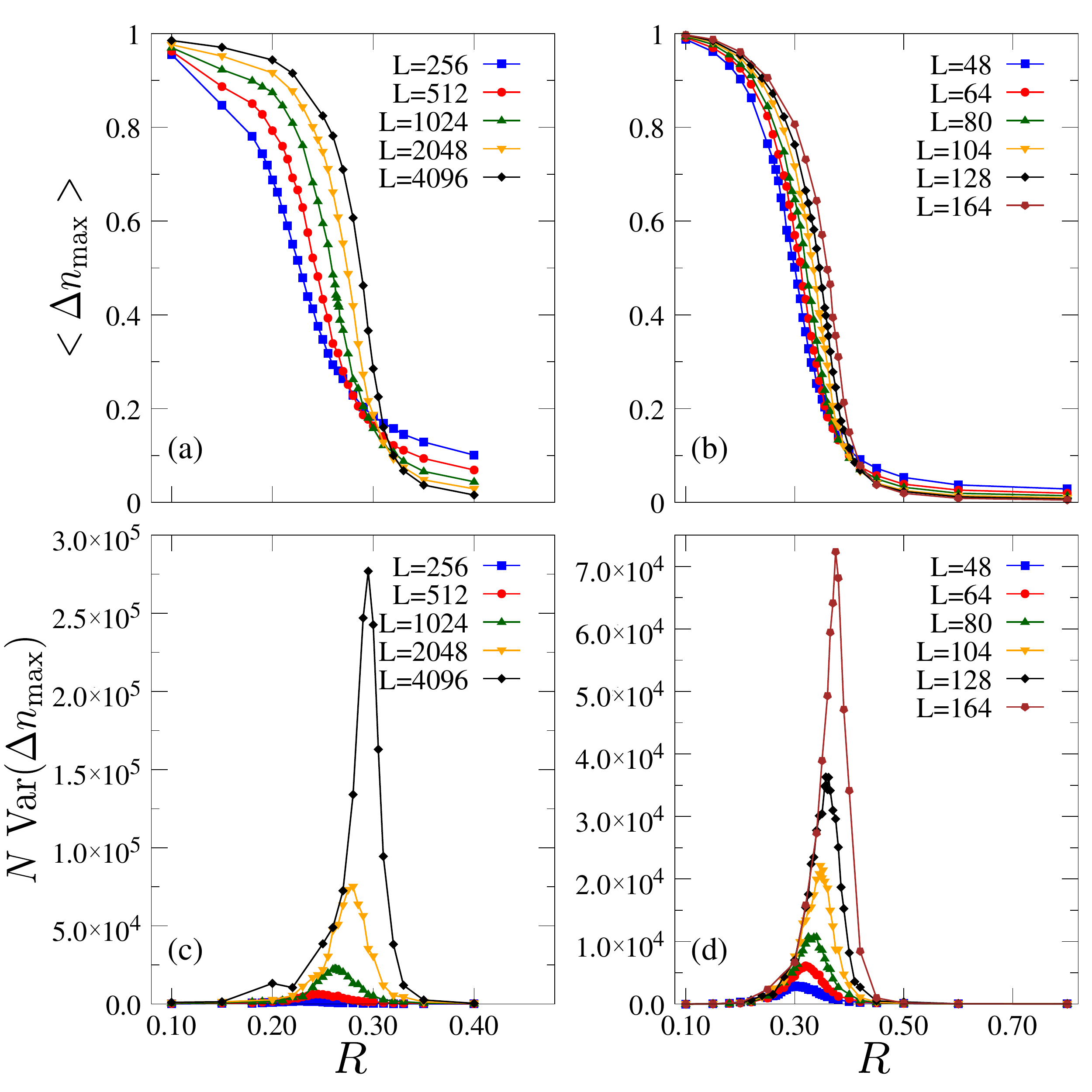}
    \caption{Evidence for a critical point in 2D and 3D EPMs. Upper: Average  value of the order parameter $\langle\Delta n_{\rm max}\rangle$ as a function of $R$ for several system sizes in 2D (a) and 3D (b). Lower: 
Variance of $\Delta n_{\rm max}$ multiplied by $N=L^D$, where $D$ is the spatial dimensions, i.e., disconnected susceptibility, in 2D (c) and 3D (d).
}
    \label{fig:critical_point}
\end{figure}

As seen in Figs.~\ref{fig:critical_point}(a,b), the average order parameter $\langle \Delta n_{\rm max} \rangle$ is small and essentially constant at high disorder strength $R$ and it starts to rapidly grow below some finite value of~$R$. Moreover,  as $L$ increases both in 2D and in 3D, the increase of $\langle \Delta n_{\rm max} \rangle$ with decreasing $R$ becomes steeper while the flat part becomes smaller, suggesting the presence of a critical point.
Figures~\ref{fig:critical_point}(c,d) show the variance of $\Delta n_{\rm max}$, which corresponds to the associated ``disconnected susceptibility", $\chi_{\rm dis}=N {\rm Var}(\Delta n_{\rm max})$, defined in analogy with an AQS driven random-field Ising model~\cite{ozawa2018random,Rossi_2022}. (It provides  crisper, but similar, results than the ``connected susceptibility" $\chi_{\rm con}=-\partial \langle \Delta n_{\rm max} \rangle /\partial\gamma$ ~\cite{ozawa2018random,ozawa2020role}.) 
The disconnected susceptibility is strongly peaked and the peak becomes sharper and higher with increasing $L$, suggesting a divergence at some critical point. Essentially the same trend is observed for $\Delta \sigma_{\rm max}$ (see SM), in agreement with the results of molecular simulations \cite{ozawa2018random}.

\begin{figure}
    \centering
    \includegraphics[width=\linewidth]{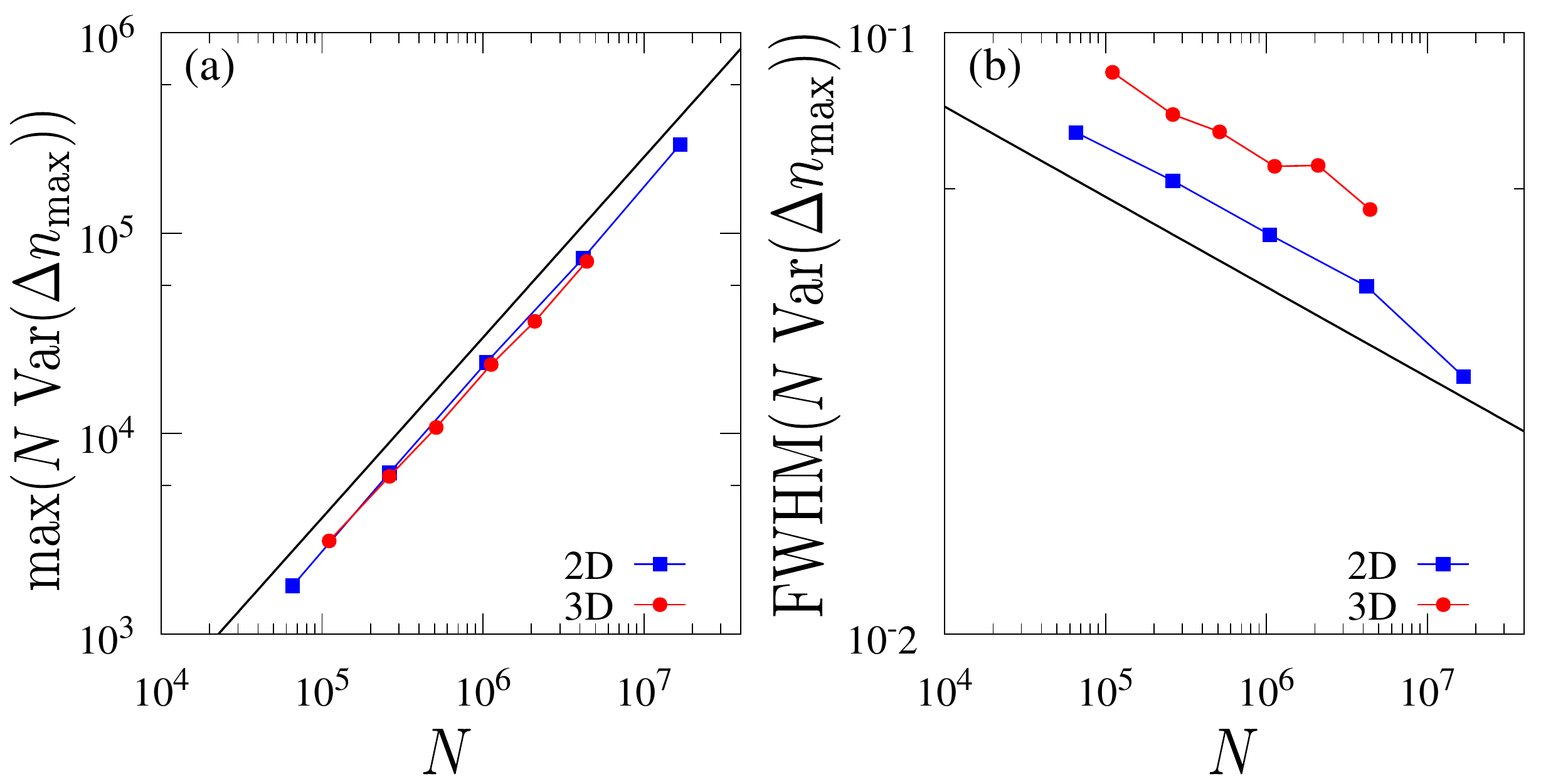}
    \caption{Log-log plot of the maximum (a) and the full width at half maximum (b) of the disconnected susceptibility associated with the order parameter $\Delta n_{\rm max}$ as a function of $N=L^D$ for both 2D (blue) and 3D (red). The straight black lines have slopes $0.9$ in (a) and $-0.15$ in (b).}
    \label{fig:FSS_max-FWHH}
\end{figure}

\begin{figure}
    \centering
    \includegraphics[width=\linewidth]{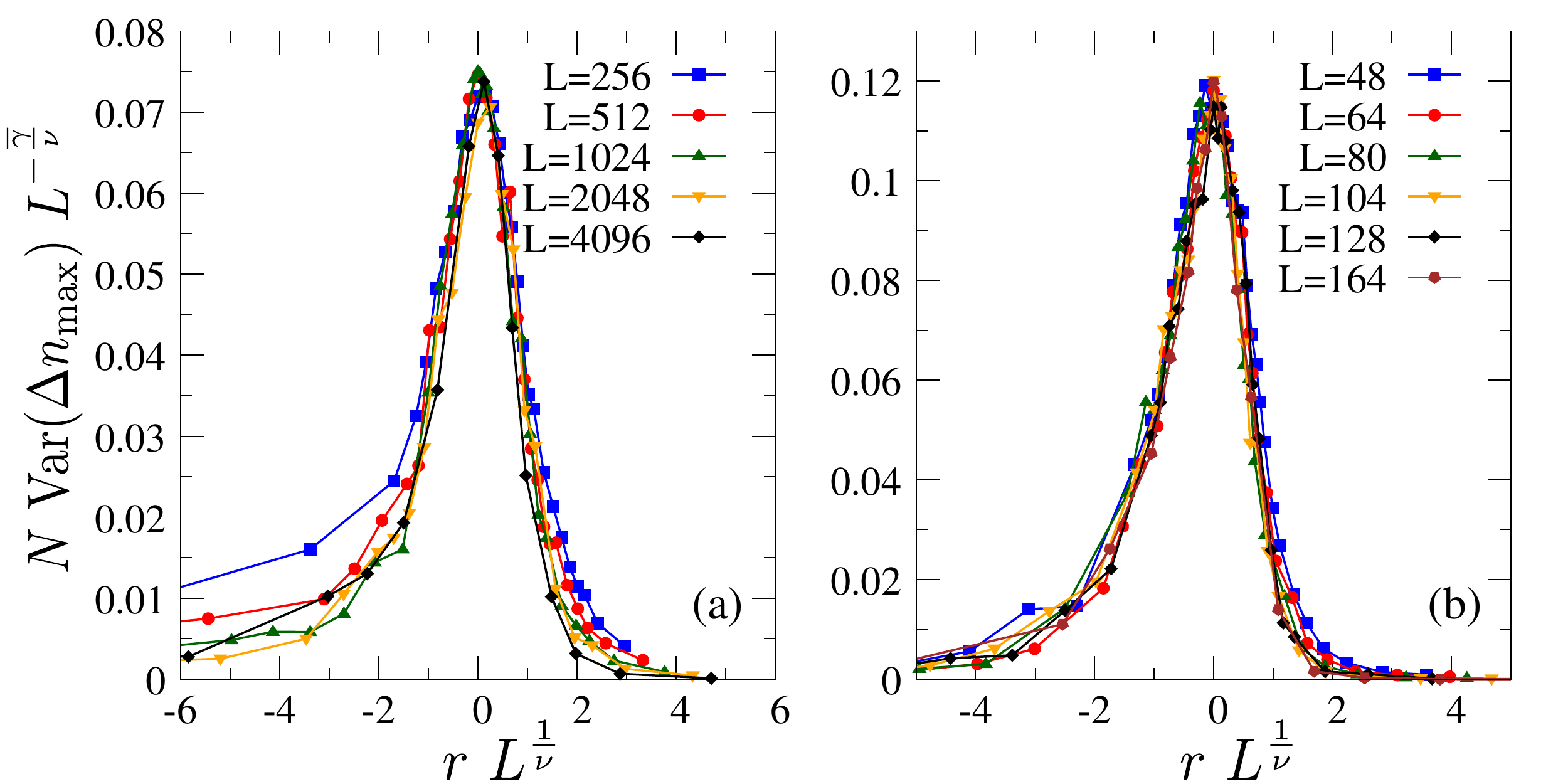}
    \caption{Scaling plot of the disconnected susceptibility versus reduced disorder $r=(R-R_{\rm c}(L))/R$ for the data in Fig.~\ref{fig:critical_point} in 2D (a) and 3D (b) EPMs.
    A good collapse is obtained for $\overline \gamma / \nu \approx 1.82$ and $\nu\approx 2.9$ in 2D and $\overline \gamma / \nu\approx 2.61$ and $\nu\approx 2.2$ in 3D. 
    }
\label{fig:scaling_disconnected}
\end{figure}

To firmly establish the existence of the critical point, we have performed a detailed finite-size scaling analysis.
We use here the same scaling ansatz as that of the AQS driven random-field Ising model, 
$\chi_{\rm dis}(r, L) \sim L^{\bar \gamma/\nu}\Psi(rL^{1/\nu})$, where
$R_{\rm c}(L)$ locates the maximum value of $\chi_{\rm dis}$,
$r=(R-R_{\rm c}(L))/R$ is the reduced disorder strength, $\Psi(\cdot)$ is a scaling function, and with $\bar \gamma$ and $\nu$ some critical exponents. According to this ansatz, the maximum over $r$ of $\chi_{\rm dis}(r, L)$ should diverge as $L^{\bar \gamma/\nu}$ and its full width at half maximum should vanish as $L^{-1/\nu}$. The corresponding plots obtained from the data in  Figs.~\ref{fig:critical_point}(c,d) are shown in Fig.~\ref{fig:FSS_max-FWHH}. We observe a good power-law behavior, and by fitting these curves we obtain $\overline\gamma/\nu = 1.86 \pm 0.02$ and $\nu = 3.0 \pm 0.3$ in 2D, and $\overline\gamma/\nu = 2.66 \pm 0.04$ and $\nu = 2.5 \pm 0.2$ in 3D, where the errors are derived from the fit. 
We also show in Fig.~\ref{fig:scaling_disconnected} the scaling collapse of the disconnected susceptibility, in which 
the parameters $\overline \gamma$, $\nu$, and $R_{\rm c}(L)$ are adjusted to provide the best visual collapse of the curves for the different values of $L$. The displayed collapses are for $\overline \gamma / \nu \approx 1.82$ and $\nu\approx 2.9$ in 2D, and $\overline \gamma / \nu\approx 2.61$ and $\nu \approx 2.2$ in 3D, values that are consistent with those  determined by the fitting procedure. 
Work is now in progress to determine whether these critical exponents are in the same universality class as an AQS driven random-field Ising model with Eshelby-like interactions~\cite{rossi_inprogress}.

Figures~\ref{fig:critical_point}, \ref{fig:FSS_max-FWHH} and \ref{fig:scaling_disconnected} provide very strong evidence for a critical behavior around $R_{\rm c}(L)$ with an estimate for the associated  exponents $\overline\gamma$ and $\nu$ in 2D and 3D. However, the critical disorder $R_{\rm c}(L)$ slightly shifts to larger $R$ as $L$ increases, as seen from  Fig.~\ref{fig:critical_point}. 
Understanding the fate of the critical disorder $R_{\rm c}(L)$ in the thermodynamic limit is therefore a key issue. 
Ref.s~\cite{ozawa2018random,ozawa2020role} proposed that $R_{\rm c}(L \to \infty)$ stays finite in the thermodynamic limit, whereas Refs.~\cite{barlow2020ductile,PhysRevResearch.4.043037,richard2021finite} argued that $R_{\rm c}(L \to \infty) \to \infty$. Note that in this second scenario there is no ductile phase for large enough system size, i.e., all systems are brittle in the thermodynamic limit (except in the singular infinite-disorder limit). We stress that the existence of a finite-disorder brittle-to-ductile critical point in the thermodynamic limit is a separate issue from the persistence of an overshoot in the average stress-versus-strain curve for large ductile systems, which was the main concern of Ref.~\cite{PhysRevResearch.4.043037}. We show below that by disentangling these two problems one can obtain conclusive evidence in favor of the existence of the critical point in the thermodynamic limit.  

\begin{figure}
    \centering
    \includegraphics[width=\linewidth]{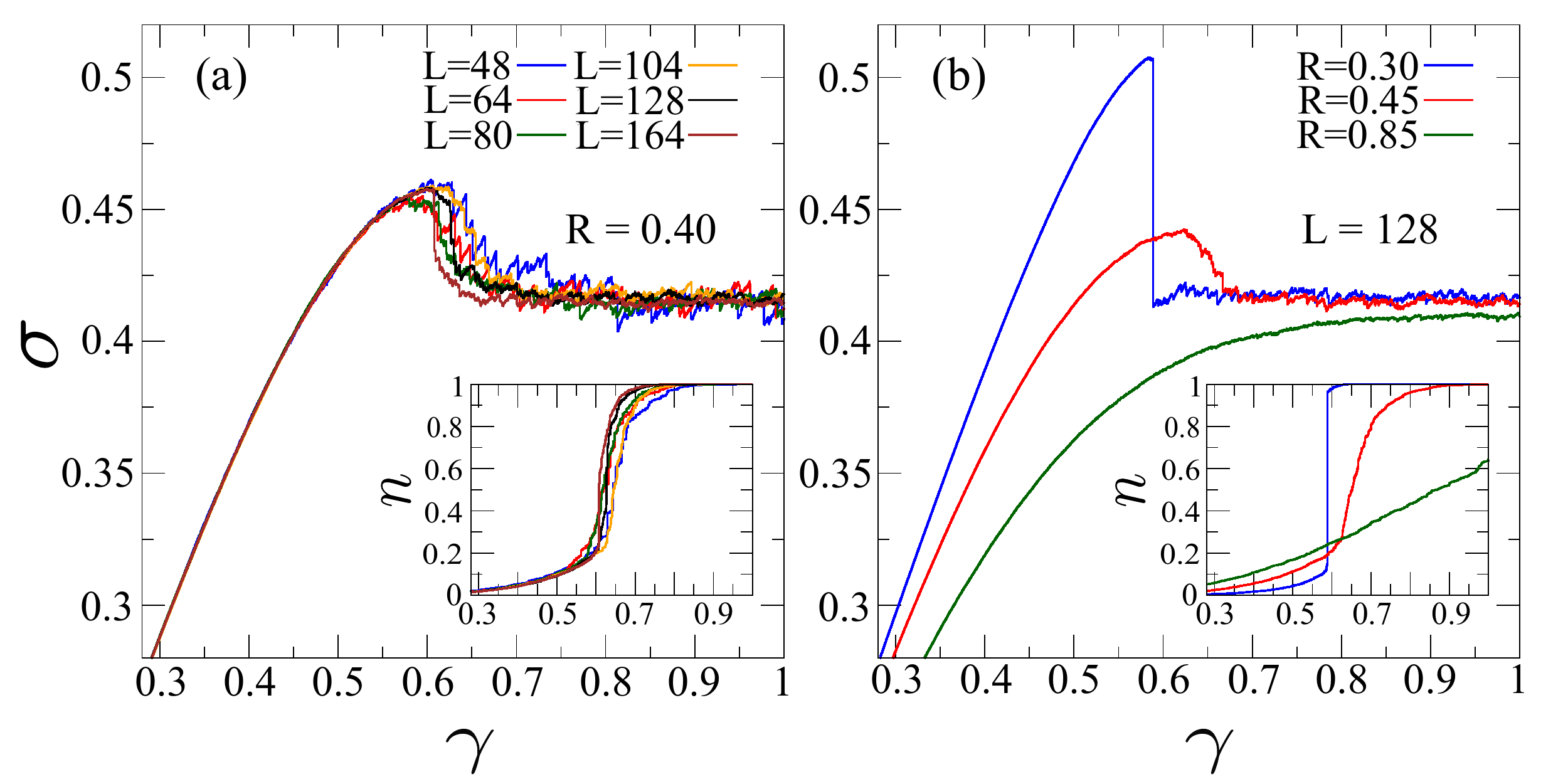}
    \caption{Stress-versus-strain curves for 3D samples at fixed disorder strength $R=0.40$ for several system sizes (a) and at fixed system size $L=128$ for different values of $R$ (b). Insets: The corresponding $n(\gamma)$ curves.}
    \label{fig:stress_vs_strain_diffR_diffL}
\end{figure}

We display in Fig.~\ref{fig:stress_vs_strain_diffR_diffL}(a) the stress-versus-strain curves of typical 3D samples at fixed $R$ for several values of $L$. We set $R=0.40$ ($>R_{\rm c}(L)$), which belongs to the putative ductile yielding regime as determined from the above finite-size scaling analysis. The plots focus on the stress values around the overshoot. 
For a fixed $R$, the stress drop tends to become sharper with increasing $L$,  showing the same trend as found in Refs.~\cite{barlow2020ductile,PhysRevResearch.4.043037,richard2021finite}.
Instead, as shown in Fig.~\ref{fig:stress_vs_strain_diffR_diffL}(b), for a fixed $L$ (here, $L=128$) a clear evolution between distinct yielding patterns is observed as $R$ is decreased, from a purely monotonic increase of the stress to a continuous overshoot and then to a discontinuous drop. 
\begin{figure}
    \centering
    \includegraphics[width=\linewidth]{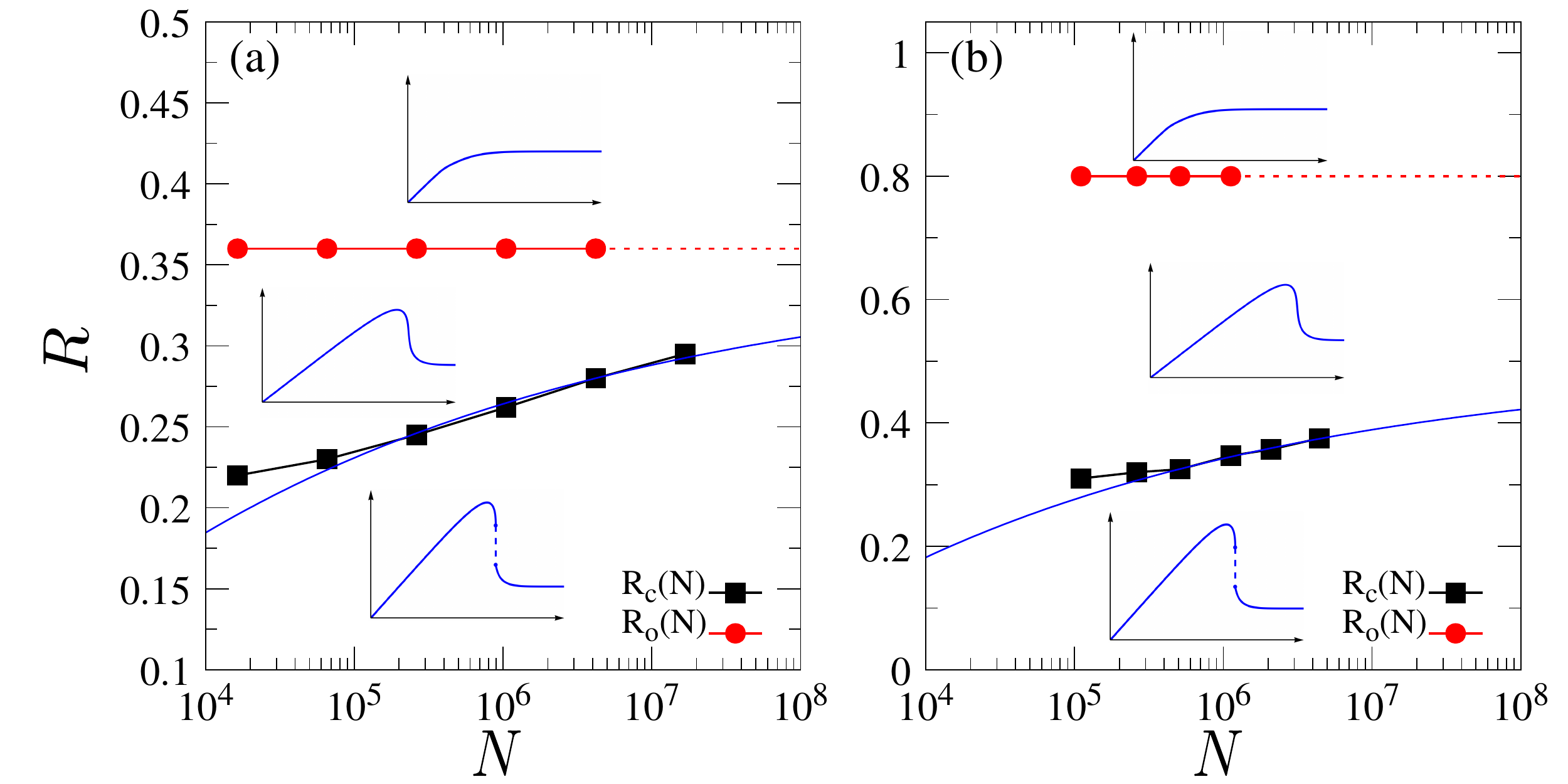}
    \caption{Value of the disorder at which the overshoot first appears, $R_{\rm o}$, and at the apparent critical point, $R_{\rm c}$, as a function of the system size, $N=L^D$, in 2D (a) and in 3D (b). 
    Blue lines are fits to $R_{\rm c}^\infty-a/N^b$, with $R_{\rm c}^\infty=0.35$, $a=0.61$, $b=0.14$ in 2D and $R_{\rm c}^\infty=0.5$, $a=1.29$, $b=0.15$ in 3D. The parameter $b$ is related to the critical exponent $\nu$ through $1/\nu=D b$, so that the fits yield  $\nu\approx 3.57$ in 2D and $\nu\approx 2.22$ in 3D. 
    Insets: The corresponding schematic stress-versus-strain curves.
    }
    \label{fig:Rc_vs_Ro}
\end{figure}
To characterize the asymptotic behavior when $L\to \infty$, we locate for each given system size $L$ the value of $R$ at which the overshoot first appears (coming from large $R$) in the average stress-versus-strain curves and we denote it by $R_{\rm o}(L)$ (see the SM for details). We display $R_{\rm o}$, together with the critical disorder $R_{\rm c}$, for 2D and 3D in Fig.~\ref{fig:Rc_vs_Ro}. To facilitate the comparison between 2D and 3D we plot $R_{\rm o}$ and $R_{\rm c}$ as a function of the number of sites $N=L^D$. We find that $R_{\rm o}$ is essentially independent of $N$ in both cases while $R_{\rm c}$ increases very slowly with $N$. As we explain in more detail below, this is direct evidence for the existence of a ductile phase over a finite range of disorder strength in the thermodynamic limit. 

The values of $R_{\rm o}(N)$ and $R_{\rm c}(N)$ define
three distinct yielding regimes in the $(N,R)$ plane, as schematically illustrated by the insets in Fig.~\ref{fig:Rc_vs_Ro}. The region $R>R_{\rm o}(N)$ corresponds to a monotonic increase of the average stress, with no overshoot. The region $R<R_{\rm c}(N)$ corresponds to a discontinuous stress drop at yielding. The regime $R_{\rm c}(N)<R<R_{\rm o}(N)$ corresponds to a continuous average stress curve with a mild overshoot. By construction, $R_{\rm c}(N)$ has to remain below $R_{\rm o}(N)$, which
then gives an upper bound on the critical disorder.
The fact that $R_{\rm o}(N)$ is essentially independent of $N$ thus provides strong evidence that $R_{\rm c}(N)$ converges to a finite value for large $N$ and that a finite-disorder brittle-to-ductile critical point persists in the thermodynamic limit. The fate of the overshoot as $N\to \infty$ is instead unclear and depends on whether $R_{\rm c}(N)$ converges to $R_{\rm o}(\infty)$ or to $R_{\rm c}(\infty)<R_{\rm o}(\infty)$ in the thermodynamic limit. In the former case the overshoot disappears at the critical point whereas a regime of ductile yielding with an overshoot exists in the latter case. We show in Fig.~\ref{fig:Rc_vs_Ro} the best fits to $R_{\rm c}(N)=R_{\rm c}(\infty)-a/N^b$ with $R_{\rm c}(\infty)$, $a$, and $b$ free parameters. We find that $R_{\rm c}(\infty)$ is finite in 2D and 3D. In the critical scaling picture and assuming that 2D and 3D are below the upper critical dimension, the parameter $b$ is related to the (correlation length) exponent $\nu$ through $1/\nu=D b$. The fits then yield $\nu\approx 3.57$ in 2D and $\nu\approx 2.22$ in 3D, values which, given the large uncertainties, are consistent with the previous determinations given above.

Strictly speaking, we cannot exclude an alternative scenario in which $R_{\rm o}(N)$ would start to increase with $N$ above some size $N^*$ which is out of reach of present-day simulations and would ultimately diverge in the thermodynamic limit together with $R_{\rm c}(N)$. However, in view of the absence of any observable $N$-dependence of $R_{\rm o}(N)$ in the accessible range, which spans three decades in 2D, and of the lack of any sound theoretical argument supporting the existence of a critical size $N^*$, this possibility seems extremely unlikely.

In conclusion, we have performed extensive numerical simulations of athermally driven elasto-plastic models in two and three dimensions. Thanks to the simple coarse-grained, lattice-based, nature of the modeling, we have been able to simulate substantially larger system sizes and larger number of samples than in molecular simulations, allowing us to perform a thorough finite-size scaling analysis.
We have obtained clear evidence for the existence of a critical point separating brittle from ductile yielding in 2D and 3D and we have provided estimates for two associated critical exponents. Our results establish, at least for the studied elasto-plastic models, that criticality persists in the thermodynamic limit and takes place for a finite value of the disorder characterizing the samples (and corresponding to a given initial stability of the solid), as suggested in Refs.~\cite{ozawa2018random,ozawa2020role}. The alternative scenario~\cite{barlow2020ductile,richard2021finite} according to which the critical point either takes place at infinite disorder or disappears because the disorder cannot go beyond some upper bound is not plausible in view of our results from elastoplastic modeling. It is nonetheless still unclear if the overshoot in the average stress-versus-strain curve disappears in the thermodynamic limit for the ductile regime, as advocated in Ref.~\cite{PhysRevResearch.4.043037}. If it does, the critical point would occur exactly when the overshoot associated with brittle yielding disappears and then gives way to a monotonic (albeit singular at criticality) average stress-strain curve. The other possibility is that the critical point 
takes place at a value of the disorder for which a smooth overshoot is still present. The latter case implies that disorder, which is not accounted for in 
the linear instability argument of Ref.~\cite{PhysRevResearch.4.043037}, is able to pin the propagation of the instability, thereby allowing for the presence of a smooth overshoot. It is hard to go beyond the present study in terms of numerical simulations. Thus, progress is now needed on the theoretical front.

Finally, we point out that the presence of a finite-disorder critical point is not restricted to the specific rheological setting considered in this paper. Recently, the AQS cyclic shear protocol has been actively studied, in relation to other nonequilibrium phase transition phenomena such as absorbing-state phase transitions. This protocol also leads to a transition from ductile to brittle-like behavior, as a function of the disorder or stability of the initial glass samples, as shown in molecular simulations~\cite{yeh2020glass,bhaumik2021role}, simulated EPMs~\cite{liu2022fate,kumar2022mapping}, as well as a mean-field EPM~\cite{parley2022mean}. A detailed characterization of the critical point under cyclic shear and the determination of the associated exponents would be an interesting subject for future research.

{\it Acknowledgments:}
We thank Ludovic Berthier, Chen Liu, Sylvain Patinet, and Matthieu Wyart for discussions.
We also thank Chen Liu for multiple advices for the numerical implementation of the elasto-plastic models. 
This work was granted access to the HPC resources of the MeSU platform at Sorbonne Université and the HPC resources of MesoPSL financed
by the Region Ile de France and the project Equip@Meso (reference
ANR-10-EQPX-29-01) of the programme Investissements d’Avenir supervised
by the Agence Nationale pour la Recherche. 
This project has received funding from the European Research Council (ERC) under the European Union's Horizon 2020 research and innovation program (grant agreement n. 723955 - GlassUniversality) and from the Simons Foundation (Grant No. 454935 to G.B.; Grant No. 454955 to F.Z.).

\bibliography{references}

\clearpage

\section{Supplemental Material}
\subsection{Implementation of Elasto-Plastic Models}

We study two and three-dimensional lattice-based elasto-plastic models~\cite{picard2004elastic} with periodic boundary conditions.
We consider a simple scalar description whose main quantity is the local shear stress $\sigma_{i}$ on each site $i$ of the lattice.
The macroscopic stress, $\sigma$ (without index), is defined by $\sigma=\frac{1}{N}\sum_i \sigma_i$.
We then assign a condition for the local stability: when a site $i$ has $|\sigma_i|> \sigma_i^{\text{th}}$ it is considered to be unstable and it yields locally with a stress drop $\eta_i>0$; otherwise this site is stable.
$\sigma_i^{\text{th}}$ is the threshold stress.
In this paper, we set it to one, $\sigma_i^{\text{th}} = 1$, uniformly for all sites for the model presented in the main text. Another choice with nonuniform $\sigma_i^{\text{th}}$'s~\cite{agoritsas2015relevance} will be discussed below. The stress drop $\eta$ is drawn from an exponential distribution, 
$P(\eta)=\frac{1}{\bar \eta}e^{-\eta/\bar \eta}$, with mean value $\bar \eta=1$.
This exponential distribution of the local stress was observed in molecular simulations~\cite{barbot2018local}.

The initial condition at $\gamma=0$,
$\sigma_{i}(\gamma=0)$, is independently and identically extracted from a probability distribution $P_0(\sigma_{i})$, given by
\begin{equation}
    P_0(\sigma_{i})= \frac{(1-\sigma_i^2)}{\mathcal{N}} e^{-\sigma_i^2/(2 R^2)}, \quad \sigma_i \in [-1,1],
\end{equation}
where $R$ is a parameter that characterizes the strength of the initial disorder of the sample and $\mathcal{N}$ is a normalization constant. 
Notice that when $R$ is small this distribution has essentially the same form as a Gaussian distribution of zero mean and standard deviation~$R$.
To ensure that each site respects the stability condition, $|\sigma_{i}(\gamma=0)|< 1$ at $\gamma=0$, we add a correction term, $1-\sigma_i^2$, in front of the Gaussian distribution. 

When the external loading is applied, the local stress at each site is increased by the same amount $\Delta \sigma_{\text{ext}}$, which is the minimum stress increment needed to induce a local instability at a single site somewhere in the sample, triggering a plastic event~\cite{vandembroucq2011mechanical}.
When this unstable site, say $i$,  yields, it influences all the other sites via a discrete stress propagator
from site $i$ to $j$, namely,
\begin{equation}
G_{j,i} = G(\textbf{r}_{j}-\textbf{r}_{i}) \ ,
\end{equation}
leading to a distribution of new stresses on sites $j\neq i$
\begin{equation}\label{eq:sigmaprop}
    \sigma_j \rightarrow \sigma_j + G_{j,i} \eta_i \ ,
\end{equation}
where $\eta_i$ is the stress drop at site $i$.
Note that choosing for convenience $G_{i,i}=-1$ allows one to use
Eq.~\eqref{eq:sigmaprop} also for $j=i$.
This stress propagation may lead to other sites reaching the condition of instability. This process goes on until all sites become stable, forming an avalanche. We next proceed to a new increase of the external stress $\Delta \sigma_{\text{ext}}$ so that a single site yields, and the process is repeated. This driving mechanism decouples the timescale of the external loading and the timescale of formation of an avalanche, which amounts to the so-called athermal quasi-static driving (AQS)~\cite{maloney2006amorphous}.

More precisely, the evolution of the model is given by the following algorithm:
\begin{enumerate}
    \item Initialize the stresses $\sigma_i(\gamma=0)$ that are i.i.d. from $P_0(\sigma_i)$;
    \item Find site $i$ that is closest to its positive threshold, i.e., minimizes $\sigma_i^{\text{th}}-\sigma_i$;
    \item Increase the stress on all sites by 
    $\Delta \sigma_{\text{ext}} = \sigma_i^{\text{th}}-\sigma_i$ such that site $i$ yields;
    \item Extract $\eta_i$ from $P(\eta)$ and evolve all sites according to $\sigma_j \rightarrow \sigma_j + G_{j,i} \eta_i$ (recall that $G_{i,i}=-1$);
    \item For all unstable sites with $|\sigma_i|>\sigma_i^{\text{th}}$, choose a random stress drop $\delta\sigma_i = \text{sign}(\sigma_i)(|\sigma_i|-\sigma_i^{\text{th}}+\eta_i)$ with $\eta_i$ i.i.d. from $P(\eta)$, assign $\delta\sigma_i=0$ to stable sites, and transform all sites in parallel according to
    $\sigma_j \rightarrow \sigma_j + \sum_i G_{j,i} \delta\sigma_i$; repeat until all sites are stable. 
    \item Repeat from (2).
\end{enumerate}
In each step, the local stresses evolve as
\begin{equation}\label{eq:sigmaiprop}
    \sigma_j \rightarrow \sigma_j + \Delta \sigma_{\text{ext}} + \sum_i G_{j,i} \Delta\sigma_i \ ,
\end{equation}
where $\Delta\sigma_i$ is the total accumulated stress change during steps 4 and 5 of the procedure.
Because
the macroscopic stress is the spatial average of $\sigma_i$, summing Eq.~\eqref{eq:sigmaiprop} over $j$ and noting that 
$\hat G_{\textbf{q}={\bf 0}}=\sum_j G_{j,i}$ does not depend on $i$,
it evolves as
\begin{equation}
    \sigma \rightarrow \sigma + \Delta \sigma_{\text{ext}} + \hat G_{\textbf{q}={\bf 0}} \frac1N\sum_{i} \Delta\sigma_i \ ,
\end{equation}
where $\textbf{q}=(q_x,q_y)$ in 2D and $\textbf{q}=(q_x,q_y, q_z)$ in 3D denote the wave-vector. 
The macroscopic strain at each step of the evolution is obtained as follows.
In the beginning, $\gamma=0$. 
At each step, the strain $\gamma$ is then increased by an amount $\Delta \gamma=\Delta \sigma_{\rm ext}/ \mu$, where $\mu$ is the local shear modulus that we set to one ($\mu=1$).
The choice of $\hat G_{\textbf{q}={\bf 0}}$ is then related to the way the system is driven. We consider in this work the strain-controlled driving, which corresponds to $\hat{G}_{{\bf q}={\bf 0}}=-1$. With this choice, the stress evolution is
\begin{equation}
\sigma \rightarrow \sigma + \mu \Delta\gamma - \Delta\sigma_p \ ,
\end{equation}
where $\Delta\sigma_p=\frac1N\sum_{i} \Delta\sigma_i$ corresponds to the irreversible plastic stress drop at constant strain.
Other driving settings including the stress-control one were for instance studied in Ref.~\cite{ferrero2019criticality}.

In this work we choose the Eshelby elastic stress propagator, which 
in the case of a continuous material is given by
\begin{eqnarray}
G^E(\textbf{r}) &=& \frac{\cos(4 \theta)}{\pi r^2} \qquad\qquad\qquad\qquad \mathrm{(2D)}, \\
G^E(\textbf{r}) &=& \frac{3}{4 \pi} \frac{r^2(x^2+y^2)-10x^2 y^2}{r^7} \qquad \mathrm{(3D)},
\end{eqnarray}
where ${\bf r}=(x,y)$ in 2D and ${\bf r}=(x,y,z)$ in 3D denote the position of a site relative to the yielded site~\cite{mura2013micromechanics}.
Notice that, since our system is defined on a discrete lattice, we need to consider a carefully discretized version of this propagator. 
First of all, we note that the value of ${G^E(\textbf{r}={\bf 0})}$ is not defined because the Eshelby kernel describes the long-range elastic field and has to be regularized at the origin. For the discrete version, we can thus freely choose $G_{i,i}=G(\textbf{r}={\bf 0})=-1$, as discussed above.
Next, we compute the Eshelby propagator in the continuous Fourier space~\cite{mura2013micromechanics}, 
\begin{equation}
\label{eqn:four_kernel}
    \begin{split}
    \hat G^E(q^2_x,q^2_y) &= - \frac{4 q_x^2 q_y^2}{\left( q_x^2 + q_y^2 \right)^2} \qquad \mathrm{(2D)}, \\
    \hat G^E(q^2_x,q^2_y,q^2_z) &= -\frac{4q_x^2 q_y^2}{q^4} -\frac{q_z^2}{q^2} \qquad \mathrm{(3D)}.
\end{split}
\end{equation}
Because we consider a finite system with a linear box length $L$ and periodic boundary conditions, we need to discretize the Fourier space as well, following the same procedure as in Ref.~\cite{ferrero2019criticality}.
First, the wave-vector components are written as 
$q_\mu = (2\pi/L) n_\mu$, with $\mu=x,y(,z)$ and $n_\mu=-L/2+1,\cdots,L/2$.
Second, we consider a discrete Laplacian instead of the continuous one, which
amounts to replacing $q^2_\mu \to 2 - 2\cos q_\mu$ in Eq.~\eqref{eqn:four_kernel}.
Moreover, Eq.~\eqref{eqn:four_kernel} is not defined at ${\bf q}={\bf 0}$ because the Eshelby propagator is not integrable in infinite volume, but in the discrete version we can normalize it in such a way that $\hat{G}_{{\bf q}={\bf 0}}=-1$, as discussed above.
 These two conditions, $\hat{G}_{\textbf{q}=\textbf{0}}=-1$ and $G_{i,i}=-1$, can be simultaneously satisfied by normalizing the Eshelby propagator in Fourier space~\cite{popovic2018elastoplastic}. Our final discrete propagator is then
 \begin{equation}
     \hat{G}_{\textbf{q}} = \begin{cases}
     -1 \ , & \textbf{q}={\bf 0} \ ,\\
     \frac{\hat G^E(q^2_\mu \to 2 - 2\cos q_\mu)}{{\cal G}} \ , & \textbf{q}=\frac{2\pi}{L}{\bf n} \neq{\bf 0} \ ,
     \end{cases}
 \end{equation}
with the constant ${\cal G}$ determined by the condition,
\begin{equation}
G_{i,i}=\frac1N \sum_{\textbf{q}} \hat{G}_{\textbf{q}}=-1 \ ,
\end{equation}
where the sum is over the discretized wave-vectors.
The propagator in discrete real space is finally obtained from
$\hat{G}_{\textbf{q}}$ by a discrete inverse Fourier transform.

\subsection{Comparison with molecular simulations} 

To perform sample averages of the various observables, we use $1000-2000$, $400-600$, $400-600$, $200-400$, and $100-200$ samples for $L=N^{1/2}=256, 512, 1024, 2048$, and $4096$, respectively, in 2D, and $800-2000$, $800-1000$, $800-1000$, $400-1000$, $200-500$, and $100-200$ samples for $L=N^{1/3}=48, 64, 80, 104, 128$, and $164$, respectively, in 3D. These numbers are larger than those used in molecular dynamics (MD) simulation studies: in previous work by some of us, we averaged over $800, 700, 400, 200, 200, 200, 200$, and $100$ samples for $N = 1000, 2000,
4000, 8000, 16 000, 32 000, 64 000$, and $128 000$, respectively, in 2D~\cite{ozawa2020role}, and $800, 400, 200, 100, 100, 50$, and $25-50$ samples for
$N = 1500, 3000, 6000, 12000, 24000, 48000$, and $96000$, respectively, in 3D~\cite{ozawa2018random}. Moreover, we study a larger number of values of the disorder ($R$ in EPMs) than in molecular simulations (in which disorder is encoded by $T_{\rm ini}$~\cite{ozawa2018random,ozawa2020role}).

It is also instructive to compare the system sizes achieved in MD and EPM simulations. This can be done by following two distinct approaches. The first one is measuring the typical size of a shear transformation zone, which corresponds to a building block (or single site) for the EPM mesoscopic description, in an MD simulation: see,  e.g.,~\cite{barbot2018local}. The second one is a quantitative calibration or mapping from MD to EPM, such that the statistical properties and the macroscopic responses in both studies match: see, e.g.,~\cite{liu2021elastoplastic,castellanos2021insights}. All of this indicates that a single site in an EPM corresponds to of the order of magnitude of 100 particles (atoms) in an MD simulation. Although these studies focus on a rather ductile yielding regime, we expect that this order of magnitude does not change significantly in the brittle yielding regime. This issue is further discussed in Ref.~\cite{zhang2022structuro}. Using this conversion, 16 millions of particles in an MD simulation as in Ref.~\cite{richard2021finite} correspond to 160000 sites in EPMs. Our largest EPM in 2D ($N=L^2=16777216$) is thus about 100 times larger and in 3D ($N=L^3=4410944$) about 30 times larger than the MD counterpart in Ref.~\cite{richard2021finite}. Therefore, effectively, our study accesses much larger system sizes than previous MD studies.


\subsection{Definition of the order parameter}

In order to characterize the presence of a critical point, we need to define a proper order parameter associated with the brittle-to-ductile transition. The most direct one is based on the presence of a discontinuity in the individual stress-versus-strain curve, namely, the stress drop, $\Delta \sigma(\gamma)$. 
One can then define $\Delta \sigma_{\rm max} = \max_{\gamma}\{\Delta \sigma(\gamma)\}$, which amounts to finding the largest stress drop in the whole dynamics~\cite{ozawa2018random,ozawa2020role}. If one considers a system with a strong disorder, i.e., in the ductile regime, this quantity goes to zero as the system size increases. However, in the brittle regime a macroscopic stress drop is present, and as a result $\Delta \sigma_{\rm max}$ remains finite. 
Therefore, one can study how $\Delta \sigma_{\rm max}$ changes when the initial stability (or disorder strength) $R$ is varied in order to distinguish the two yielding regimes. 

\begin{figure}
    \centering
    \includegraphics[width=\linewidth]{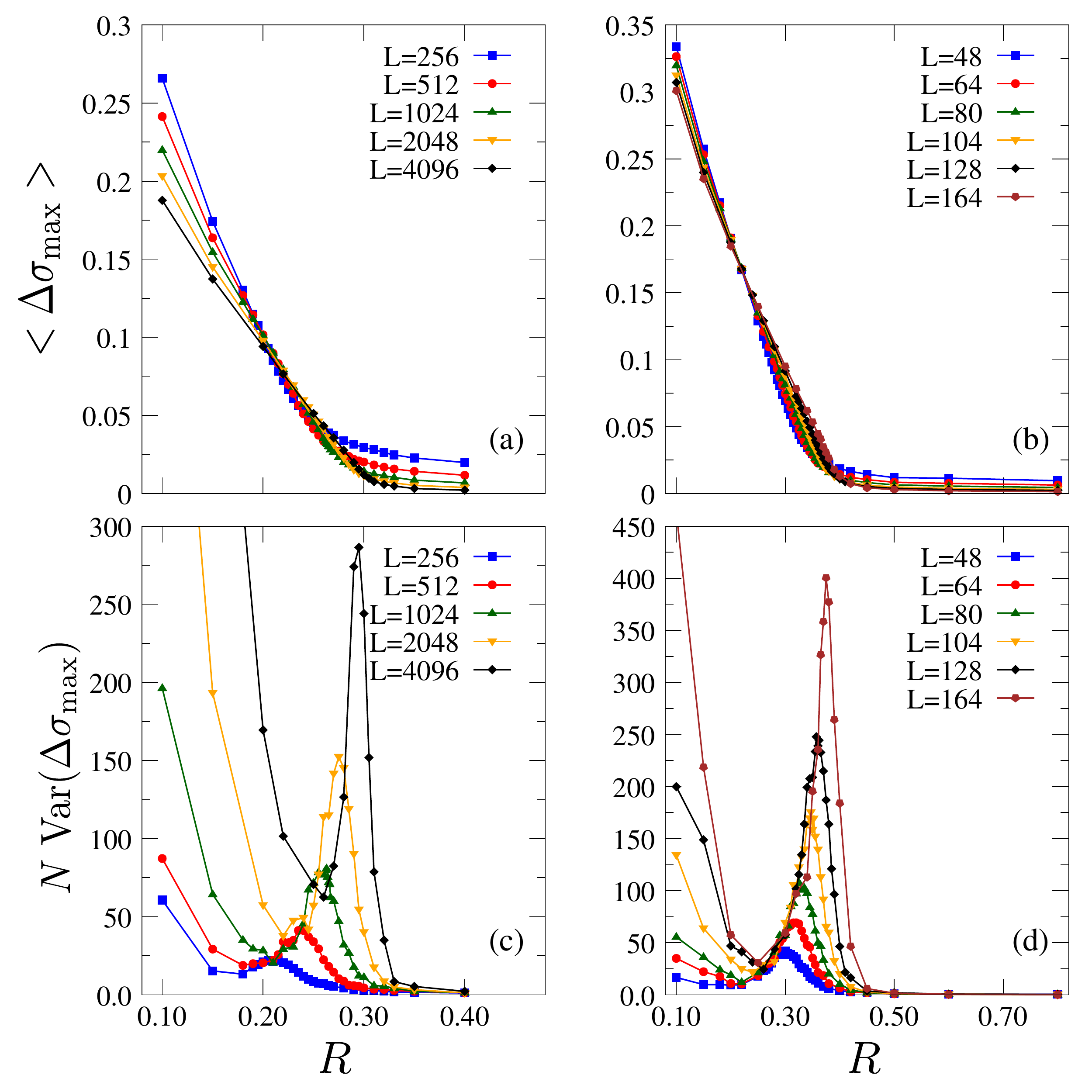}
    \caption{Mean value of the order parameter $\Delta \sigma_{\rm max}$ as a function of $R$ for several system sizes $L$ in 2D (a) and 3D (b).
    The corresponding variance of $\Delta \sigma_{\rm max}$, which corresponds to a disconnected susceptibility, is shown in (c) for 2D and (d) for 3D.}
    \label{fig:critical_point_delta_sigma}
\end{figure}

The mean and variance of $\Delta \sigma_{\rm max}$ are presented in Fig.~\ref{fig:critical_point_delta_sigma} in both 2D and 3D.
As in Fig.~2 in the main text, we observe the signature of a critical point: $\langle \Delta \sigma_{\rm max} \rangle$ is of order O(1) when $R$ is small, while it goes to nearly zero when $R$ is large, and more so as the system size $L$ increases. Around the onset of the growth of $\langle \Delta \sigma_{\rm max} \rangle$, the (disconnected) susceptibility, $N\left( \langle \Delta \sigma_{\rm max}^2 \rangle - \langle \Delta \sigma_{\rm max} \rangle^2 \right)$, shows a peak that grows with the system size. However, as we decrease $R$ further, the susceptibility seems to diverge again with the system size. This unexpected behavior can be explained by considering a simple argument in the case of very weak disorder $R$. Let us assume that when $R \to 0^+$ the system evolves purely elastically up to the macroscopic yielding strain $\gamma_{\rm Y}$, which corresponds to the value of $\gamma$ at which the least stable site yields. We now define the maximum stress of the initial distribution, $\sigma_{\rm max}=\max_i\{ \sigma_i(\gamma=0) \}$. Consequently, we have $\mu \gamma_{\rm Y} = \sigma^{\rm th}-\sigma_{\rm max}$. If we also assume that the macroscopic stress goes to a plateau value $\sigma_{\rm ss}$ after yielding, we obtain $\Delta \sigma_{\rm max}= \mu \gamma_Y-\sigma_{\rm ss}$. Therefore, we can relate the fluctuations of $\Delta \sigma_{\rm max}$ and $\sigma_{\rm max}$ through
\begin{equation}
    \langle \Delta \sigma^2_{\rm max} \rangle- \langle \Delta \sigma_{\rm max} \rangle^2 = \langle \sigma_{\rm max}^2\rangle- \langle\sigma_{\rm max}\rangle^2 \ .
\end{equation}
Since the initial distribution is nearly a Gaussian, $\sigma_{\rm max}$ follows a Gumbel distribution, associated with extreme-value statistics~\cite{fortin2015applications}. We then arrive at
\begin{equation}
    \langle \Delta \sigma^2_{\rm max} \rangle-\langle\Delta \sigma_{\rm max}\rangle^2 = \frac{\pi^2 R^2}{12 \ln N} \ .
\end{equation}
This expression explains the growth of the susceptibility at $R \to 0$ in Fig.~\ref{fig:critical_point_delta_sigma}.
In real experiments and molecular simulations, such a growth would not be present because in practice the elastic branch contains multiple plastic events and the purely elastic assumption taken above does not hold in these systems~\cite{lemaitre2021anomalous}.

In order to avoid this problem in the elasto-plastic models, we have decided to consider a different order parameter. Since the presence of a macroscopic stress drop in the brittle regime is accompanied by the abrupt formation of a shear band, one can think of characterizing yielding by looking at the number of sites that yield in each horizontal or vertical line (in 2D) or plane (in 3D). More precisely, we define a quantity $n_{x,y}(\gamma)$ in 2D that is $1$ if the site at $(x,y)$ has already had a plastic event during its evolution up to $\gamma$, and $0$ otherwise (and similarly for $n_{x,y,z}(\gamma)$ in 3D). We can then define 
\begin{align}
        n_x(\gamma)&=\frac{1}{L}\sum_{y=1}^L n_{x,y}(\gamma) \qquad \mathrm{(2D)}, \\ 
        n_x(\gamma)&=\frac{1}{L^2}\sum_{y,z=1}^L n_{x,y,z}(\gamma) \qquad \mathrm{(3D)},  
\end{align}
in 2D and 3D respectively. In the same way we also define $n_y(\gamma)$ for vertical lines in 2D, and $n_y(\gamma)$ and $n_z(\gamma)$ for vertical planes in 3D. We expect that, in the ductile regime, there is essentially no difference between $n_x(\gamma)$ and $n_y(\gamma)$ (and $n_z(\gamma)$) since the plastic activity occurs rather uniformly throughout the sample and no spatial organization of the events is observed. On the other hand, in the brittle regime, there is a specific $x^*$ (or $y^*$ or $z^*$) such that $n_{x^*}(\gamma) \approx 1$ at $\gamma \approx \gamma_Y$ due to the formation of the shear band. This corresponds to a strong anisotropic localization of the plastic activity. One can then characterize whether the system is in a ductile or brittle phase by studying the evolution of these quantities. In particular we consider $n(\gamma)=\max_{x,y} \{n_x(\gamma),n_y(\gamma) \}$ (and similarly in 3D). The inset of
Fig.~4 in the main text
shows the evolution of $n(\gamma)$ for some samples. The discontinuous stress drop of $\sigma(\gamma)$ in the brittle regime corresponds to the discontinuous jump of $n(\gamma)$, whereas the mild continuous crossover of $\sigma(\gamma)$ in the ductile regime corresponds to a continuous increase of $n(\gamma)$. Therefore, $n(\gamma)$ essentially contains the same information as $\sigma(\gamma)$ as far as characterizing brittle and ductile yielding is concerned.
As for $\Delta \sigma(\gamma)$, we compute the jumps, $\Delta n(\gamma)$, in this quantity, which allows us to define a new order parameter $\Delta n_{\rm max}$, defined as $\Delta n_{\rm max} = \max_{\gamma}\{\Delta n(\gamma)\}$. The results for $\Delta n_{\rm max}(\gamma)$ are presented in Fig.~2 in the main text.
\\


\subsection{Force balance}

Since we use the AQS driving, imposing force balance is an important element in the simulation of the EPM. 
In the continuum limit,
mechanical equilibrium imposes that the divergence of the stress vanishes. For example in 2D, one should have

\begin{equation}
\frac{\partial\sigma_{\rm xx}(x,y)}{\partial x} +  \frac{\partial\sigma_{\rm xy}(x,y)}{\partial y}=0 \ .
\end{equation}
Integrating this relation over $x$, we obtain
\begin{equation}\label{S17}
\int dx \frac{\partial\sigma_{\rm xy}(x,y)}{\partial y}
=
-\int dx 
\frac{\partial\sigma_{\rm xx}(x,y)}{\partial x} \ .
\end{equation}
Because the right-hand side of 
Eq.~\eqref{S17} vanishes due to periodic boundary conditions for any $y$, we obtain
\begin{equation}
\frac{\partial}{\partial y} \int dx\, \sigma_{\rm xy}(x,y) =0 \ ,
\end{equation}
implying that the average shear stress $\sigma_{\rm xy}(x,y)$ must be constant over a row (or a column). A similar derivation can be obtained in 3D.

In the discrete version, averaging the local shear stress (here $\sigma=\sigma_{\rm xy}$) evolution in Eq.~\eqref{eq:sigmaiprop} over $x$ in 2D, 
we obtain
\begin{equation}\begin{split}
    \frac1L\sum_x \sigma_{x,y} \rightarrow &
    \frac1L\sum_x \sigma_{x,y} + \Delta\sigma_{\rm ext}
    -\Delta\sigma_p \\ &
    +\frac1N \sum_{q_y\neq 0}\hat G_{q_x=0,q_y} \Delta \hat \sigma_{q_x=0,q_y} \ e^{i q_y y} \ .
\end{split}\end{equation}
Because the Eshelby kernel in Fourier space vanishes whenever $q_x=0$ for all $q_y \neq 0$, see Eq.~\eqref{eqn:four_kernel}, the second line vanishes and
the average stress along all rows changes by the same quantity, $\Delta\sigma_{\rm ext} - \Delta\sigma_p$~\cite{liu2016critical}. 
The same result is obtained for columns. This means that the difference between the average stress along a given row and along a different row or column remains constant during the dynamics.    
Therefore, to impose mechanical equilibrium, or force balance condition, it is sufficient to ensure that the sum of the stresses along rows and columns is the same for the initial stress distribution at $\gamma=0$.
For 3D, a similar argument holds for the average local shear stress over an x-z (or y-z) plane.

As already explained, we initialize the local stresses by a nearly Gaussian distribution with zero mean. Thus, the force balance condition is satisfied asymptotically at large $L$ because the average stress over a single row or column goes to zero when $\gamma=0$ and $L\to\infty$.
However, this condition does not hold exactly in finite system sizes.
We have thus tested another way to initialize the local stresses, which strictly enforces the force balance condition. The initial stress at each site is drawn from a Gaussian distribution of zero mean and standard deviation $R$,
but we follow the procedure described in Ref.~\cite{popovic2018elastoplastic} and implement an additional random operation to keep the sum of the stresses equal along all rows and columns. We then use this initial condition for our EPMs, and we find that the results quickly converge to the results presented in the main text with increasing $L$. Therefore, our conclusions on the critical point are not affected by the choice of the force balance condition at $\gamma=0$.

\begin{figure}[t]
    \centering
    \includegraphics[width=\linewidth]{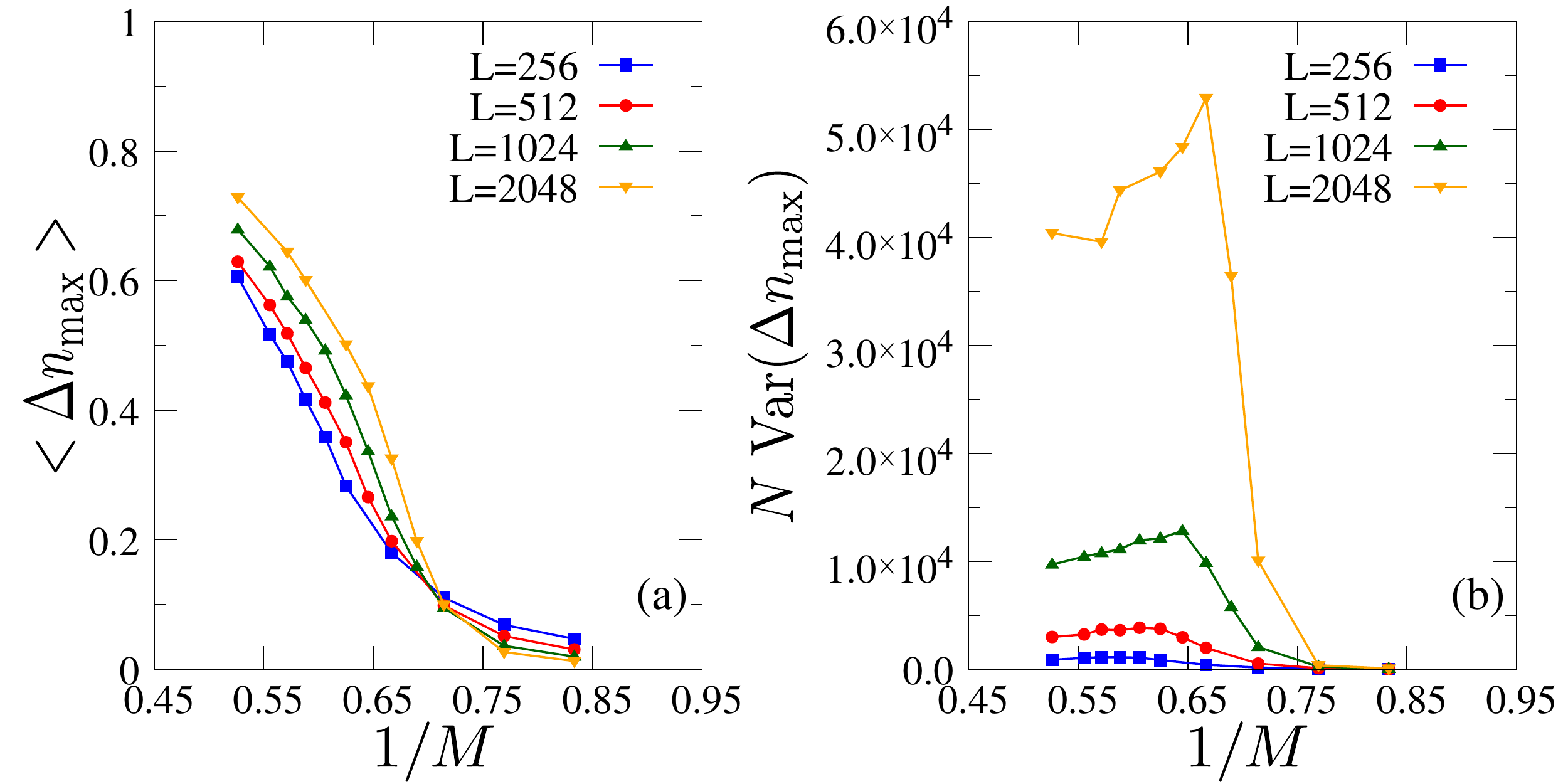}
    \caption{Finite-size analysis of the critical point for the model described in Ref.~\cite{popovic2018elastoplastic}. (a) Average value of $\Delta n_{\rm max}$. (b) Associated disconnected susceptibility.}
    \label{fig:critical_point_wyartmodel}
\end{figure}

\begin{figure}[t]
    \centering
    \includegraphics[width=\linewidth]{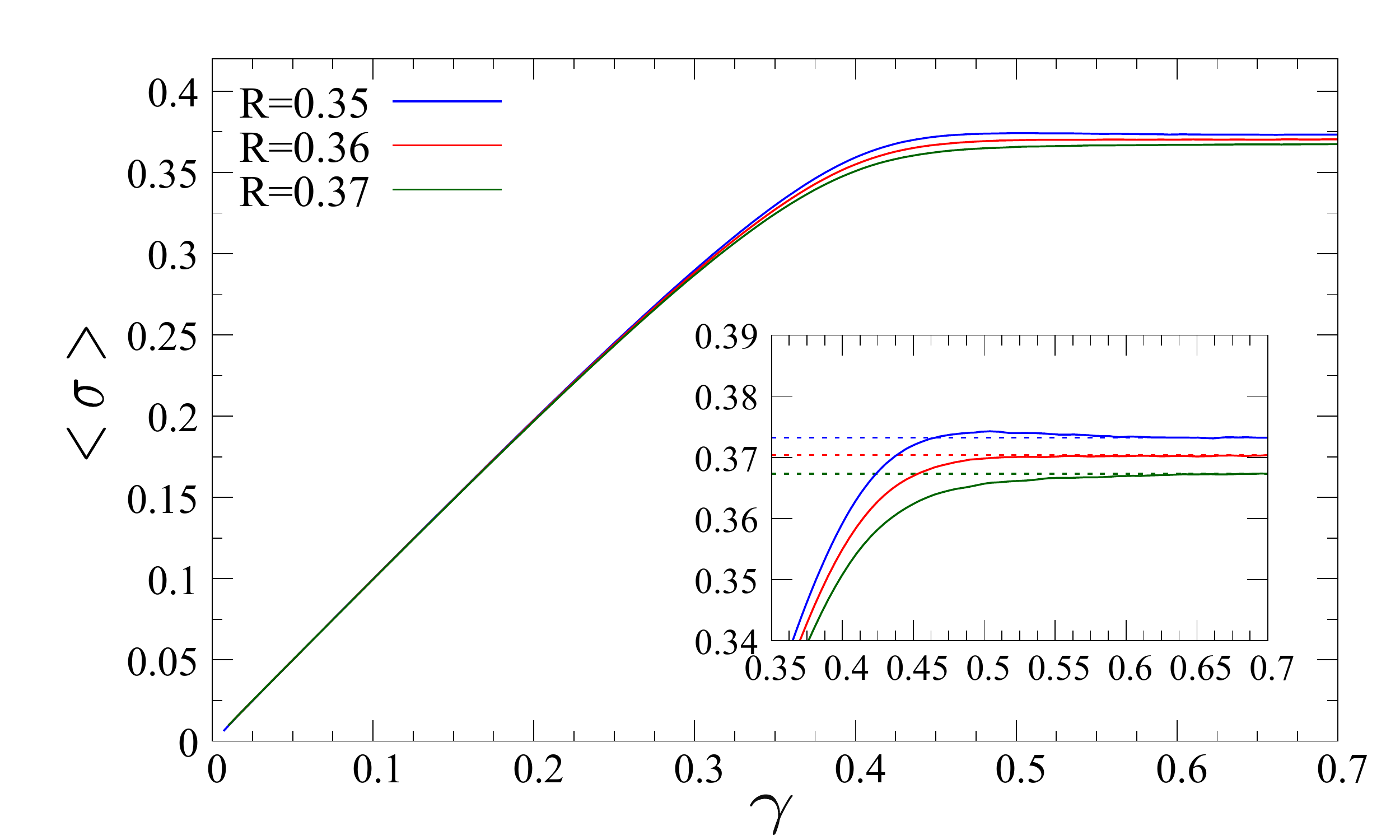}
    \caption{Average stress-versus-strain curves for a 2D model with $L=2048$ close to $R_{\rm o}=0.36$. Inset: zoom in on the region where the overshoot appears. The dashed horizontal lines are a guide for the eye.}
    \label{fig:overshoot_stress_vs_strain}
\end{figure}

\subsection{Variations of the model: random thresholds}

We have also considered a different way to tune the disorder of the sample. We fix $R=0.45$ for the initial condition and we let instead the stress threshold $\sigma_i^{\rm th}$ be randomly distributed~\cite{popovic2018elastoplastic}. In this model, the local stress threshold  at each site, $\sigma_i^{\rm th}$, when $\gamma=0$ is chosen according to a Gaussian distribution of mean $M$ and variance $0.01$. After each plastic event, $\sigma_i^{\rm th}$ is updated from another Gaussian distribution of mean $1$ and variance $0.01$. Notice that in this model the system shows brittle or ductile yielding depending on the average value $M$ of the initial $\sigma_i^{\rm th}$. A large $M$ corresponds to a stable or weakly disordered sample, whereas a small $M$ corresponds to a less stable or highly disordered sample. 
Although this model contains some ingredients that are different than those of  the model we have studied in the main text, such as the threshold distribution and its softening, the precise way to enforce force balance, etc., it essentially displays the same phenomenology. This is illustrated in Fig.~\ref{fig:critical_point_wyartmodel} for the mean value of the order parameter $\Delta n_{\rm max}$ and the associated disconnected susceptibility,  with $1/M$ playing the role of $R$. 
These numerical observations suggest that the critical point that we have  identified in this work is quite robust and universal with respect to changes in the details of the simulated elasto-plastic models.


\subsection{Determination of the onset of the stress overshoot}

We determine the onset value $R_{\rm o}$ at which the stress overshoot appears (or disappears) in the evolution with the applied strain of the stress  averaged over many independent samples. We illustrate the operational procedure in Fig.~\ref{fig:overshoot_stress_vs_strain} for the 2D elasto-plastic model with $L=2048$. Three averaged stress-versus-strain curves in the vicinity of the first appearance of the overshoot are shown. From such a plot, one can detect $R_{\rm o}$ in a reliable manner. We have performed this analysis for different values of $L$ both in 2D and 3D. The outcome is summarized in Fig.~5 in the main text.

\end{document}